\documentstyle[aps,prb,multicol,epsf]{revtex}
\begin{document}
\draft
\title{Vortices in magnetically coupled superconducting layered systems}
\author{Roman G. Mints}
\address{School of Physics and Astronomy, Raymond
and Beverly Sackler Faculty of Exact Sciences,\\
Tel Aviv University, Tel Aviv 69978, Israel}
\author{Vladimir G. Kogan and John R. Clem}
\address{Ames Laboratory - DOE and Department of Physics and Astronomy,
Iowa State University, Ames Iowa 50011}
\date{\today}
\maketitle
\begin{abstract}
Pancake vortices in stacks of thin superconducting films or layers are
considered. It is stressed that in the absence of Josephson coupling
topological restrictions upon possible configurations of vortices are
removed and various examples of structures forbidden in bulk
superconductors are given. In particular, it is shown that vortices may
skip surface layers in samples of less than a certain size $R_c$ which
might be macroscopic. The Josephson coupling suppresses $R_c$
estimates.
\end{abstract}
\begin{multicols}{2}
\narrowtext
\section{Introduction}
Attempts to measure the anisotropy parameter in Bi- and Tl-based
high-$T_c$ superconducting compounds yield very large values
$\sim 100 - 300$.\cite{gamma} In some layered organic superconductors
this parameter is even higher.\cite{organic} This has led many to
believe that the Josephson coupling in these materials is so weak that
for many purposes it can be disregarded altogether. Models of vortices
in these compounds based on pure magnetic coupling between
two-dimensional pancake vortices proved to suffice for many
applications.\cite{Efetov,Guinea,Buzdin,Art,Clem,Blatter}
\par
We consider in this paper a system of thin superconducting layers
coupled only via the magnetic field between them. This system is
qualitatively different from the bulk superconductors: there is no phase
coherence across the layers. In three-dimensional bulk materials
vortices are banned from terminating inside superconductor because of
topology: the phase changes by $2\pi$ when one circles the vortex core
at which the phase is singular. A core termination would have meant that
the phase could acquire the $2\pi$ change along a path that does not
contain a singularity.
\par
Clearly, this ban does not hold in layered systems with no Josephson
coupling. A vortex perpendicular to the layers or, better to say, a
stack of pancake vortices (or ``pancakes"), may terminate,\footnote{We 
use the term ``vortex termination in the layer N'' for the situation when
the vortex cores are present in the layer N and in all layers under,
whereas the cores are absent in the layers above N up to the the sample
surface.} in principle, at any layer and channel the flux at least
partially into interlayer space. This event might be energetically
unlikely, but it is not forbidden by topology of the phase, which is
defined separately in each layer.  \par We argue below that within the
model of layered materials with zero Josephson coupling such termination
in layers adjacent to the surface might be energetically favorable in
samples of finite size. We use the formal technique suggested by one of
us,\cite{Kogan} which is reviewed briefly. The technique allows us to
compare energies of various configurations in a straightforward manner. As
one of the applications, we compare energies of two configurations: one
with a vortex piercing all layers of a half-space multilayer and another
with the vortex terminating in one of the top layers. We show that the
energy cost of the subsurface vortex termination diverges with the sample
size $R$: \begin{equation}
\Delta\epsilon \sim {\phi_0^2\over 16\pi^2\Lambda}
\ln\frac{R}{\kappa\lambda_{ab}}\,
\label{eq1}
\end{equation}
where $\Lambda =\lambda_{ab}^2/s$, $\lambda_{ab}$ is the
penetration depth, $s$ is the layer periodicity,
$\kappa =\lambda_{ab}/\xi_{ab}$, and $\xi_{ab}$ is the
coherence length. The divergence of $\Delta\epsilon$ is weak, however,
and in samples smaller than $R_c\sim\kappa\lambda_{ab}$
the subsurface termination becomes energetically
favorable. With $\lambda_{ab}\approx 0.3\,\mu$m, $\kappa\approx 100$
for Bi-2212, we estimate $R_c\sim 30\,\mu$m  with a large numerical
factor  so that $R_c$ may reach a macroscopic size.
\par
This estimate is reduced considerably by including the Josephson
coupling because one has to include in the balance the  energy of
Josephson strings channelling the flux sideways into the interlayer
space. We estimate, however, that even then $R_c$ remains on the order
of microns. This suggests that if the surface of a layered compound like
Bi-2212 has defects separated by $R_c$, the occurrence of vortex cores
at the surface is a rare event. We also show that for a vortex
terminating at a depth $\ll\lambda_{ab}$  under the surface, most of the
magnetic flux crosses the core-free layers into the free space, the
absence of the core notwithstanding. Hence, we expect vortices
terminating under the surface to be invisible for the scanning tunneling
technique, but detectable in decoration experiments.
\par
Other examples of possible but unusual configurations of pancake
vortices in layered materials are considered. It is shown that a vortex
in a film of a {\it finite} size and containing several superconducting
layers may have normal cores only in internal layers and carry flux
different from the flux quantum $\phi_0$. The last feature appears also
in exotic configurations such as a stack of pancake vortices situated in
every other layer. Implications of these possibilities are discussed.
\par
\section{Approach}
We begin with a brief review of thin superconducting films. As was
stressed by Pearl,\cite{Pearl1} the situation in a thin film differs
from that of a bulk since a large contribution to the energy of a vortex
comes from the  stray fields. In fact, the problem of a vortex in a thin
film is reduced to that of the field  distribution in free space subject
to certain boundary conditions at the film surface. Since
${\rm curl}\,{\bf h}={\rm div}\,{\bf h}=0$ outside, one can introduce a
scalar potential for the {\it outside} field:
\begin{equation}
{\bf h}=\nabla \varphi\,,\qquad \nabla^2\varphi =0\,.
\label{eq2}
\end{equation}
To formulate the boundary conditions for the outside Laplace problem let
us consider a film of thickness $d\ll \lambda $ occupying the $xy$
plane; $\lambda $ is the bulk penetration depth of the film material.
For a vortex at ${\bf r}=0$, the London equations for the film interior
read:
\begin{equation}
{\bf h}+{4\pi\lambda^2\over c}\,{\rm curl}\,{\bf j}=
\phi_0 {\hat {\bf z}}\,\delta ({\bf r})\,,
\label{eq3}
\end{equation}
where ${\hat {\bf z}}$ is the unit vector along the vortex axis.
Averaging over the thickness $d$ we obtain
\begin{equation}
h_z+{4\pi\Lambda\over c}\,{\rm curl}_z {\bf g}=
\phi_0 \delta ({\bf r} )\,,
\label{eq4}
\end{equation}
where ${\bf g}({\bf r})$ is the sheet current density, ${\bf r}=(x,y)$,
and $\Lambda=\lambda^2/d$ is the film penetration depth. Other
components of London equation turn identities after averaging.
\par
Since all derivatives $\partial/\partial z$ are large relative to the
tangential $\partial/\partial{\bf r}$, the Maxwell equation
${\rm curl}\,{\bf h}=4\pi{\bf j}/c$  is reduced to conditions relating
the sheet current to discontinuities of  the tangential field:
\begin{equation}
{4\pi\over c} g_x=h_y^--h_y^+\,,\quad
{4\pi\over c} g_y=h_x^+-h_x^-\,.
\label{eq5}
\end{equation}
Here the superscripts $\pm $ stand for the upper and lower ($z=\pm d/2$)
faces of the film. The field component perpendicular to the film, $h_z$,
is the same at both film faces.
\par
Substituting Eq. (\ref{eq5}) into Eq. (\ref{eq4}) and using
div$\,{\bf h}=0$, we obtain:
\begin{equation}
h_z + \Lambda\,\Big(\frac{\partial h_z^-}{\partial z}-
\frac{\partial h_z^+}{\partial z}\Big) =\phi_0 \delta ({\bf r} )\,.
\label{eq6}
\end{equation}
This equation along with
\begin{equation}
 h_z^+ = h_z^-\,,
\label{eq7}
\end{equation}
and conditions at infinity constitute the boundary conditions for the
Laplace problem, Eq. (\ref{eq2}), of the field distribution outside the
film.
\par
Let us turn to the question of energy. We consider a general situation
of vortices in {\it finite bulk} samples. The energy consists of the
London energy (magnetic + kinetic) inside the sample, $\epsilon^{(i)}$,
and the magnetic energy outside, $\epsilon^{(a)}$:
\begin{equation}
\epsilon^{(i)} =\int {dV\over 8\pi}\Big[h^2+\Big({4\pi\lambda
\over c}\Big)^2j^2\Big],\quad \epsilon^{(a)}=\int {dV\over 8\pi}\,h^2.
\label{eq8}
\end{equation}
Then, for the potential gauged to  zero at $\infty$ (which is possible
in zero applied field) one has
\begin{equation}
8\pi \epsilon^{(a)}=\oint\varphi{\bf h}\cdot d{\bf S}\,,
\label{eq9}
\end{equation}
where the integral is over the sample's surface with $d{\bf S}$ directed
inward the material. If (as it may happen in some situations considered
below) $\varphi$ does not vanish at infinity, one has to examine the
integral (\ref{eq9}) at infinity.
\par
The London part can be transformed integrating the  kinetic term by
parts:
\begin{equation}
8\pi \epsilon^{(i)}={4\pi\lambda^2\over c}\oint({\bf h}\times {\bf j})
\cdot d{\bf S}\,,
\label{eq10}
\end{equation}
where the integral is over the samples surface and the surface of the
vortex core; there might be more than one sample in the system, while
not all of them may contain vortices. The integral over the samples
surface is further transformed:
$$\oint d{\bf S}\cdot ({\bf j}\times\nabla\varphi)=
\oint d{\bf S}\cdot\varphi (\nabla \times {\bf j})$$
(take a closed contour at the sample surface, consider the total sample
surface as made of two pieces supported by this contour, and apply the
Stokes theorem to the integration over each piece to show that
$\oint d{\bf S}\cdot(\nabla\times \varphi{\bf j})=0$).
Combining the result with $\epsilon^{(a)}$ of (\ref{eq9}), one obtains
$$\oint d{\bf S} \cdot \varphi
\Big({\bf h}+{4\pi\lambda^2\over c}\,{\rm curl}{\bf j}\Big)\,.$$
The expression in parentheses is zero for samples with no vortices,
whereas for those containing vortices it is
$\phi_0 \hat{{\bf v}}\delta^{(2)}({\bf r}-{\bf r}_v)$ where
$\hat{{\bf v}}$ is the direction of the vortex crossing the surface at
the point ${\bf r}_v$ ($\delta^{(2)}({\bf r}-{\bf r}_v)$ is the
two-dimensional $\delta$ function). We then obtain
\begin{equation}
\frac{8\pi}{\phi_0}\,\epsilon =\varphi({\bf r}_{ent})-
\varphi({\bf r}_{ex})-\frac{4\pi\lambda^2}{\phi_0c}
\oint_{core} d{\bf S}\cdot({\bf h}\times {\bf j})\,,
\label{eq11}
\end{equation}
with ${\bf r}_{ent}$ and ${\bf r}_{ex}$ being the positions
of the vortex entry and exit at the sample surface (the vortex is
assumed to cross the sample surface at right angles). Note that if
there are more than one superconductor present, but the vortex pierces
only one of them, the result (\ref{eq11}) still holds (although
$\varphi$'s differ for each particular configuration). For more than
one vortex in the system, one has to sum up over all vortices. For
thin films, the integral over the core surface ($\propto d$) can be
neglected in Eq. (\ref{eq11}).
\par
It is instructive to see now how the known Pearl results
\cite{Pearl1,Pearl2} for vortex in a film and a bulk half-space can be
obtained within the approach outlined here. We also demonstrate an added
advantage of the method, a relatively simple way to evaluate energies.
\par
\subsection{Vortex in a thin film}
Consider a thin film situated at $z=0$. The general form of the
potential which vanishes at $z\rightarrow +\infty$ of the empty
upper half-space is
\begin{equation}
\varphi ({\bf r},z>0)=\int\frac{d^2{\bf k}}{(2\pi)^2}\,
\varphi_u({\bf k})\,e^{i{\bf k}\cdot{\bf r}-kz}\,.
\label{eq12}
\end{equation}
with $k=\sqrt{k_x^2+k_y^2}$. In the lower half-space we have to replace
$z\rightarrow -z$ in Eq. (\ref{eq12}). The two dimensional (2D) Fourier
transforms $\varphi_u({\bf k})$ and $\varphi_l({\bf k})$ for the upper
and lower half-spaces are obtained with the help of boundary conditions
(\ref{eq6},\ref{eq7}):
\begin{equation}
-k\varphi_u +\Lambda k^2(\varphi_l-\varphi_u ) = \phi_0\,, \qquad
-k\varphi_u = k\varphi_l\,.
\label{eq13}
\end{equation}
This system yields:
\begin{equation}
\varphi_u =-\frac{\phi_0}{k(1+2k\Lambda)}\,.
\label{eq14}
\end{equation}
\par
We point out first that the total flux crossing a plane $z=\,{\rm const}$
is
\begin{equation}
\Phi_z=\int h_z\,d^2{\bf r} = h_z({\bf k}=0)=
- k\,\varphi_u e^{-kz}|_{k=0}=\phi_0
\label{eq15}
\end{equation}
for any $z$, i.e., the film is crossed by the flux $\phi_0$. And the
second: according to (\ref{eq11}), the energy $\epsilon$ of the Pearl
vortex is given by
\begin{eqnarray}
{8\pi\epsilon\over\phi_0}
&=&\varphi_l(0)-\varphi_u(0)=-2\varphi_u(0)
=\int\frac{d^2{\bf k}}{(2\pi)^2}\,\frac{2\phi_0}{k(1+2k\Lambda)}
\nonumber\\
&=&{\phi_0\over \pi}\int_0^{2\pi/\xi}\frac{dk}{1+2k\Lambda}\,,
\label{eq16}
\end{eqnarray}
where the cutoff at $k_{max}\approx 2\pi/\xi$ is introduced to a
logarithmically divergent integral. This yields
\begin{equation}
\epsilon=\frac{\phi_0^2}{16\pi^2\Lambda}
\ln\Big(4\pi{\Lambda\over\xi}\Big)\,.
\label{eq17}
\end{equation}
\par
\subsection{Vortex in a half-space}
Let now the half-space $z<0$ be filled with a superconductor having the
penetration depth $\lambda $. The stray field  in the free space $z>0$
is given by the potential (\ref{eq12}). Within the superconductor we
have the London equation (\ref{eq3}). The general solution can be
written as
\begin{equation}
{\bf h}={\bf h}^{(0)}+{\bf h}^{(v)},
\label{eq18}
\end{equation}
where ${\bf h}^{(0)}$ solves  Eq. (\ref{eq3}) with zero right-hand side
(RHS), whereas ${\bf h}^{(v)}$ is a particular solution of the full
Eq. (\ref{eq3}). The latter can be taken as the field of an infinitely
long unperturbed vortex along $z$; this assures correct singular
behavior at the vortex axis. The Fourier transform of this field is
\begin{equation}
{\bf h}^{(v)}=\frac{\phi_0}{1+\lambda^2k^2 }\, {\hat{\bf z}}\,.
\label{eq19}
\end{equation}
\par
We now Fourier transform the homogeneous Eq.(\ref{eq3}) for
$h^{(0)}_i({\bf r})$ with respect to the variable ${\bf r}$ and obtain
\begin{equation}
h_i^{(0)}({\bf k},z)(1+k^2\lambda^2)
-\lambda^2\partial^2_zh_i^{(0)}({\bf k},z)=0\,,
\label{eq20}
\end{equation}
where $i=x,y,z.$ The solution which vanishes deep in the superconductor
is
\begin{equation}
h_i^{(0)}({\bf k},z)=H_i({\bf k})\,e^{q z}\,,\qquad
q=\sqrt{1+\lambda^2k^2}/\lambda\,.
\label{eq21}
\end{equation}
Here, $H_i({\bf k})$'s are not independent: ${\rm div}\,{\bf h}^{(0)}=0$
gives 
\begin{equation}
ik_xH_x+ik_yH_y+q\,H_z=0\,.
\label{eq22}
\end{equation}
\par
The boundary conditions at $z=0$ read:
\begin{eqnarray}
ik_{x,y}\,\varphi_u &=&H_{x,y} \,,
\label{eq23}\\
-k \,\varphi_u &=&H_z +h_z^{(v)}\,.
\label{eq24}
\end{eqnarray}
The four Eqs. (\ref{eq22}-\ref{eq24}) suffice to determine the four
unknowns, $\varphi_u$ and $H_{x,y,z}$.  We obtain
\begin{equation}
\varphi =-\frac{\phi_0 }{kq(k +q)\lambda ^2}\,,\qquad
H_z = -\frac{\phi_0k}{q^2(q +k)\lambda^2}\,.
\label{eq25}
\end{equation}
We do not write down $H_{x,y}$ which describe the spreading of the
vortex field under the surface. Equations (\ref{eq25}) coincide with
Pearl's solution.\cite{Pearl2}
\par
To evaluate the energy of the vortex in this case with the help of
the general result (\ref{eq11}), we first calculate the potential at
the vortex exit:
\begin{equation}
\varphi (0)=\int\frac{d^2{\bf k}}{(2\pi)^2}\,\varphi({\bf k})=
-\frac{\phi_0}{2\pi\lambda}\,.
\label{eq26}
\end{equation}
The energy associated with the core, i.e., the integral over the core
surface in Eq. (\ref{eq11}), is
\begin{equation}
\frac{2c}{\lambda^2}\,\epsilon_c =
2\pi\xi\int_{-\infty}^0 dz [h_zj_{\theta}]_{r=\xi}\,.
\label{eq27}
\end{equation}
Integrating the London relation
\begin{equation}
j_{\theta}=-\frac{c\phi_0}{2\pi^2\lambda^2}\,\Big(\nabla\theta+
{2\pi\over \phi_0}\,{\bf A}\Big)_{\theta}
\label{eq28}
\end{equation}
over a circle of a radius $r$, we obtain for the current
density near the core
\begin{equation}
j_{\theta}=\frac{c\phi_0}{8\pi^2\lambda^2\,r}
\label{eq29}
\end{equation}
which  results in
\begin{equation}
\epsilon_c =\frac{\phi_0}{8\pi }\int_{-\infty}^0 dz\,h_z\,.
\label{eq30}
\end{equation}
Using $H_z$ of Eq. (\ref{eq25}) with Eqs. (\ref{eq11}) and
(\ref{eq26}) we obtain the energy cost of vortex termination at the
surface as
\begin{equation}
\frac{\phi_0^2}{16\pi^2\lambda}\Big({\pi\over 2}-1\Big)
\approx\frac{0.57}{\ln\kappa}\,\epsilon_L\lambda\,,
\label{eq31}
\end{equation}
where $\epsilon_L$ is the energy per unit length of an unperturbed
vortex.
\par
\section{Subsurface termination}
\subsection{Half-space $+$ a thin layer}
We now turn to a stack of pancake vortices perpendicular to the layers
of a half-space layered sample. The layered structure has the period
$s$ whereas each layer is characterized by the Pearl length
$\Lambda=\lambda^2/d$ with $d$ being the layer thickness.  Note that the
{\it measurable} length $\lambda_{ab}$ is related to other lengths
in the problem by
\begin{equation}
\lambda_{ab}^2=\lambda^2\,{s\over d}=\Lambda\,s\,.
\label{eq32}
\end{equation}
\par
Let us consider first the stack with a missing pancake in the top layer.
We model the rest of the stack as a continuous half-space $z<0$ with the
penetration depth $\lambda_{ab}$.\cite{Clem} A thin film with the
Pearl length $\Lambda$ is situated  at $z=s$. In the two domains of
free space, $z>s$ and $0<z<s$, the 2D Fourier transforms of the
potential are:
\begin{eqnarray}
\varphi_u({\bf k})\,e^{-kz},\qquad\qquad\qquad z>s\,,
\label{eq33} \\
\varphi_1({\bf k})\,e^{kz}+\varphi_2({\bf k})\,e^{-kz}\,,\quad 0<z<s\,.
\label{eq34}
\end{eqnarray}
On the film at $z=s$, we have the boundary conditions (\ref{eq7}) and
(\ref{eq6}).
Three more equations are provided by the field continuity at $z=0$. The
requirement ${\rm div}\,{\bf h}^{(0)}=0$, Eq. (\ref{eq22}), completes
the system thus providing enough equations for
$\varphi_u,\varphi_{1,2}$ and $H_{x,y,z}$:
\begin{eqnarray}
\varphi_u
&=&\frac{2\phi_0\Lambda e^{2ks}}{Dq\lambda_{ab}^2}\,,
\qquad\varphi_1 =\frac{\phi_0 }{Dqk\lambda_{ab}^2}\,,
\nonumber\\
\varphi_2
&=&\frac{\phi_0 e^{2ks}}{Dqk\lambda_{ab}^2}\,(1+2k\Lambda )\,,
\label{eq35}\\
D&=&q -k-(1+2k\Lambda )(k+q)e^{2ks}\,\nonumber
\end{eqnarray}
and
\begin{equation}
H_z({\bf k})= \frac{\phi_0}{Dq\lambda_{ab}^2}\,
\Big[1-e^{2ks}(1+2k\Lambda )\Big] -
\frac{\phi_0}{q^2\lambda_{ab}^2}\,,
\label{eq36}
\end{equation}
where $q$ is defined in Eq. (\ref{eq21}); $H_{x,y}$ is not needed
here.
\par
The fluxes through planes $z={\rm const}$ are:
\begin{eqnarray}
\Phi_z =\,\phi_0\,\frac{\Lambda}{\Lambda +\lambda_{ab}}\,,
\quad\quad z>s\,,
\label{eq37}\\
\Phi_z  = \phi_0 \,,\qquad\qquad\qquad z<0\,.
\label{eq38}
\end{eqnarray}
Within the layer $0<z<s$, the flux $\Phi_z$ decreases from the bulk
value $\phi_0$ to the value (\ref{eq37}). Therefore, a small fraction
of the flux,
\begin{equation}
\phi_0 \lambda_{ab}/\Lambda =\phi_0 s/\lambda_{ab}\,,
\label{eq39}
\end{equation}
is channelled aside between the half-space and the top film.
\par
Using the same formal scheme, we can solve the problem when the film
at $z=s$ contains a pancake vortex. The only difference with the case
of ``half-space + an empty film" is in the London boundary condition
(\ref{eq7}) on the film at $z=s$. We skip details and provide the
result denoting the new solutions with a star:
\begin{eqnarray}
\varphi^*_u
&=& \varphi_u+\frac{\phi_0 e^{ks}}{kD}[k-q+(k+q)e^{2ks}]\,,
\nonumber\\
\varphi^*_{1,2}
&=&\varphi_{1,2}\mp\frac{\phi_0 e^{ks}}{D k}\,(q\pm k)\,,
\label{eq40}\\
H^*_z&=&H_z-{2k\phi_0\over D}\,.\nonumber
\end{eqnarray}
\par
The energy cost of vortex termination, i.e., the difference between the
situations without and with the top pancake, is given by
\FL
\begin{eqnarray}
{8\pi\over\phi_0}\Delta\epsilon
&=&[-\varphi_{ex}(0)]-
[\varphi^*_{ent}(s)-\varphi^*_{ex}(s) -\varphi^*_{ex}(0)]
\nonumber\\
&+& \int_{-\infty}^0 dz\,(h_z-h_z^*);
\label{eq41}
\end{eqnarray}
the last line  is the difference in core contributions
of vortex lines in the low half-space, see Eq. (\ref{eq30})
\par
Evaluating the Fourier integrals for $\varphi$'s, one notes that the
integration over $k$ is done within the domain $ 2\pi/R\,, 2\pi/\xi$
where $R$ is the sample size and $\xi$ is the coherence length (usually
called $\xi_{ab}$). For $s < \xi$ one can expand $e^{-2ks}$ in the
denominator $D$ and see that the term with $s$ can be neglected. Indeed,
by introducing a new variable $\sinh u=k\lambda_{ab}$ and utilizing
$\Lambda \gg\lambda_{ab}$, we obtain
\begin{equation}
D\approx -2k[1+(\Lambda/\lambda_{ab})e^u]\approx
-2ke^u\Lambda/\lambda_{ab}.
\label{eq42}
\end{equation}
This results in a factor $1/\Lambda = s/\lambda_{ab}^2$ in front of
the integrals. Then, in the linear approximation in $s$, we can set
$s=0$ in the integrand after which all integrals become simple. We
obtain after a straightforward algebra:
\begin{equation}
\Delta\epsilon={\phi_0^2\over 16\pi^2\Lambda}
\ln\Big[\frac{R}{(4\pi)^{3/2}\kappa\lambda_{ab}}\,\Big].
\label{eq43}
\end{equation}
\par
It is of interest to obtain the current distribution in the top film.
To this end, we write Eq. (\ref{eq12}) in cylindric coordinates:
\begin{equation}
\varphi ({\bf r})=\int\frac{d k\,k}{2\pi }\,J_0(kr)\,
\varphi({\bf k})\,e^{-kz}\,.
\label{eq44}
\end{equation}
The tangential fields above and under the film are:
\begin{eqnarray}
h_r^+
&=&-\int_0^{\infty}{dk\over 2\pi}\,k^2J_1(kr)\,\varphi_u\,e^{-ks}\,,
\label{eq45}\\
h_r^-
&=&-\int_0^{\infty}{dk\over 2\pi}\,k^2J_1(kr)\,
(\varphi_1e^{ks}+\varphi_2e^{-ks})\,.
\nonumber
\end{eqnarray}
The sheet current is given by
\begin{equation}
{4\pi\over c} g_{\varphi}=h_r^+-h_r^-=
2\int_0^{\infty}{k^2dk\over 2\pi}\,J_1(kr)\,\varphi_1e^{ks}\,,
\label{eq46}
\end{equation}
where the boundary condition (\ref{eq7}) at $z=s$,
$\varphi_u+\varphi_1e^{2ks}-\varphi_2 = 0$, had been used. With the
help of Eq. (\ref{eq42}) for $D$, we obtain after integration (see
Ref.\onlinecite{Grad}, 6.663):
\begin{equation}
g_{\varphi}=\frac{c\phi_0}{8\pi^2\Lambda\lambda_{ab}}\,
I_1\Big({r\over 2\lambda_{ab}}\Big)
\,K_0\Big({r\over 2\lambda_{ab}}\Big)\,
\label{eq47}
\end{equation}
where $I_1,K_0$ are Modified Bessel functions. Thus, $g_{\varphi}$
vanishes as $r\ln (2\lambda_{ab}/r)$ when $r\rightarrow 0$, and as
$1/r$ for $r\gg \lambda_{ab}$.
\par
We note that the same formal scheme can be applied to consider
the termination of the pancake stack in the second, third, etc, layer
from the top. The length $\Lambda$ characterizing the top layers
with no cores should then be taken as $\Lambda/2$, $\Lambda/3$,
etc.
\par
\subsection{Vortex in a finite stack of layers}
As an example of such a system we consider a pancake in middle layer of
a ``short stack'' of three films. Let the pancake sit at the origin of 
the film at $z=0$; the film is characterized by the length $\Lambda$. At
$z\pm s$ two other films are situated having the film penetration depth
$\Lambda_1 \ne \Lambda$. Following the same method we write the 2D Fourier
transforms: 
\FL
\begin{eqnarray}
\qquad&& \varphi_a \,e^{-kz}\,, \qquad\qquad\qquad z>s\,,
\nonumber\\
\qquad&& \varphi_{b1} \,e^{kz}+\varphi_{b2}\,e^{-kz}\,,
\quad 0<z<s\,,
\label{eq48}\\
\qquad&& \varphi_{c1}\,e^{kz}+\varphi_{c2}\,e^{-kz}\,,
\quad -s<z<0\,.
\nonumber
\end{eqnarray}
Due to the symmetry, $h_{bz}(z)=h_{cz}(-z)$, and
$\varphi_{c1}=-\varphi_{b2}$,
$\varphi_{c2}=-\varphi_{b1}$; this leaves only three
coefficients $\varphi$ to be determined. The boundary conditions
(\ref{eq7}) and (\ref{eq6}) at $z=s$ give
\begin{eqnarray}
\varphi_a  +\varphi_{b1} \,e^{2ks}-\varphi_{b2} =0\,,
\label{eq49}\\
\varphi_a(1+k\Lambda_1)-
k\Lambda_1(\varphi_{b1}\,e^{2ks}+\varphi_{b2})=0\,.
\label{eq50}
\end{eqnarray}
The third equation is provided by the London boundary condition
(\ref{eq6}) at $z=0$:
\begin{equation}
\varphi_{b1}(1-2k\Lambda)-\varphi_{b2}(1+2k\Lambda)=\phi_0/k\,.
\label{eq51}
\end{equation}
The continuity condition (\ref{eq7}) for $h_z$ at $z=0$ is satisfied
identically. Equations (\ref{eq49}-\ref{eq51}) yield
\begin{eqnarray}
\varphi_a&=&\frac{2\phi_0 \Lambda_1}{D_1} \,,\qquad
\varphi_{b1}=\frac{\phi_0 }{kD_1}\,e^{-2ks}
\nonumber\\
\varphi_{b2}&=&\frac{\phi_0}{kD_1}\,(1+2k\Lambda_1)\,,
\label{eq52}\\
D_1&=&(1-2k\Lambda) e^{-2ks} -(1+2k\Lambda) (1+2k\Lambda_1)\,.
\nonumber
\end{eqnarray}
\par
The fluxes are
\begin{eqnarray}
\Phi_z(z>s) &=& \phi_0\frac{\Lambda_1}{2\Lambda+\Lambda_1+s}
\nonumber\\
\Phi_z(0<z<s) &=& \phi_0\frac{\Lambda_1+s-z}{2\Lambda+\Lambda_1+s}\,.
\label{eq53}
\end{eqnarray}
Note that the flux $\phi_0s/(2\Lambda+\Lambda_1)\ll \phi_0$ is deflected
into the space between the layers.  For $\Lambda \gg \Lambda_1 \gg s$,
\begin{equation}
\Phi_z = \phi_0\frac{\Lambda_1}{2\Lambda }\,.
\label{eq54}
\end{equation}
Note also that for $\Lambda_1=\Lambda\gg s$, the flux piercing the
sandwich made of identical films is $\phi_0/3$. The energy is given by
\FL
\begin{eqnarray}
&&{8\pi\epsilon\over\phi_0}=
[\varphi_c -\varphi_b]_{{\bf r}=z=0}
=-2\int\frac{d^2{\bf k}}{(2\pi)^2}\,
(\varphi_{b1} +\varphi_{b2})\,.
\label{eq55}
\end{eqnarray}
\par
Using the same argument as that leading to Eq. (\ref{eq42}), we obtain
for the denominator
\begin{equation}
D_1\approx
-2k\Lambda\Big(\frac{2\Lambda +\Lambda_1}{\Lambda}+2k\Lambda_1\Big)\,.
\label{eq56}
\end{equation}
Integration in Eq. (\ref{eq55}) is now easy.
\par
Consider the case $\Lambda_1=\Lambda$ and set $s=0$ (everywhere
except in $\Lambda =\lambda^2/s$). Integration yields for the energy
of one pancake in a stack of three films
\begin{equation}
\epsilon_{010} \approx
\frac{\phi_0^2/3}{16\pi^2\Lambda}\ln\frac{16\pi^2\Lambda^2R}{9\xi^3}\,.
\label{eq57}
\end{equation}
where it was assumed that the sample size $R\gg 2\pi\Lambda$ and the
subscript $010$ is to indicate that the pancake is situated in the
middle film. This is to be compared with a vortex piercing all three
layers, which can be considered as a Pearl vortex in a single film with
the effective film penetration depth $\Lambda /3$. Then, according to
Eq. (\ref{eq17}),
\begin{equation}
\epsilon_{111}=
\frac{3\phi_0^2}{16\pi^2\Lambda}
\ln\frac{4\pi\Lambda}{3\xi}\,,
\label{eq58}
\end{equation}
and the subscript $111$ indicates that each layer has a pancake. The
difference of energies (\ref{eq57}) and (\ref{eq58}) is negative if
\begin{equation}
R<\Lambda (\Lambda/\xi)^6
\label{eq59}
\end{equation}
and we have omitted a large numerical factor. This suggests that
in the stack of three films, vortices carrying the flux $\phi_0/3$ with
a core only in the middle film might be energetically preferable to
standard vortices piercing the whole stack with the flux $\phi_0$. 
\par
It is worth noting that because our solution holds for
$\Lambda\ne\Lambda_1$, in fact, it applies also to a vortex piercing $N$
middle layers of the $2N_1+N$ stack of layers provided that both $N_1s$
and $Ns$ are less than $\lambda_{ab}$; this allows us to treat the
$N_1s$ as one film with $\Lambda_1=\lambda_{ab}^2/N_1s$.
\par
\section{Infinite dilute stack of pancakes}
We consider now an infinite stack of layers. Let the layer at $z=0$ and
all even layers be free of pancakes; the pancakes are situated at
$z=\pm s$ and at all odd films (at ${\bf r}=0$). Consider three regions:
(a) $s<z<2s$, (b) $0<z<s$, and (c) $-s<z<0$.
\par
\epsfclipon 
\begin{figure} 
\narrowtext 
\epsfxsize=0.7\hsize 
\centerline{ 
\vbox{ 
\epsffile{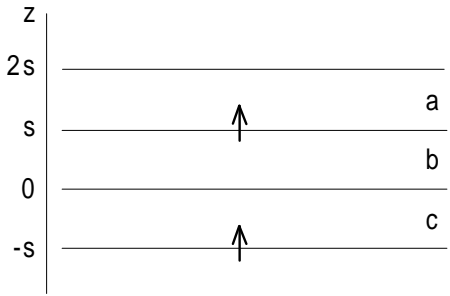}
\caption{The unit cell of a periodic structure $0<z<2s$.}
}} 
\label{f1} 
\end{figure} 
The 2D Fourier transforms:
\begin{eqnarray}
\varphi_a=\varphi_{a1}\,e^{kz}+\varphi_{a2} \,e^{-kz}\,,
\label{eq60}\\
\varphi_b=\varphi_{b1}\,e^{kz}+\varphi_{b2} \,e^{-kz}\,,
\label{eq61}\\
\varphi_c=\varphi_{c1}\,e^{kz}+\varphi_{c2} \,e^{-kz}\,.
\label{eq62}
\end{eqnarray}
Due to the symmetry with respect to $z=0$,
$h_{bz}(z)=h_{cz}(-z)$. This gives
$\varphi_{c1}=-\varphi_{b2}$,
$\varphi_{c2}=-\varphi_{b1}$.
Further relations are provided by periodicity which can be expressed
as $h_{az}(z)=h_{cz}(z-2s)$ and yields
\begin{equation}
\varphi_{a1} =-\varphi_{b2}\,e^{-2ks}\,,\qquad
\varphi_{a2}= -\,\varphi_{b1} \,e^{2ks}\,.
\label{eq63}
\end{equation}
Thus, out of 6 unknown coefficients of the system (\ref{eq60}-\ref{eq62})
we are left with only two. As the two needed equation one can take the
London conditions (\ref{eq6}) at $z=0$ and $z=s$:
\begin{eqnarray}
\varphi_{b1}(1-2k\Lambda)-\varphi_{b2}(1+2k\Lambda)
&=& 0\,,
\nonumber\\
\varphi_{b1}e^{ks}(1+2k\Lambda)-\varphi_{b2}e^{-ks}(1-2k\Lambda)
&=& \phi_0/k\,.
\label{eq64}
\end{eqnarray}
This gives
\begin{eqnarray}
\varphi_{b1}&=&\frac{\phi_0 }{kD_2}(1+2k\Lambda)\,,
\nonumber\\
\varphi_{b2}&=&\frac{\phi_0 }{kD_2}(1-2k\Lambda)\,,
\label{eq65}\\
D_2&=&e^{ks} (1+2k\Lambda)^2 -e^{-ks} (1-2k\Lambda)^2\,.
\nonumber
\end{eqnarray}
\par
The flux at any plane $z=$const within the domain (b) is readily
evaluated:
\begin{eqnarray}
\Phi_b(z)
&=& h_{bz}(k=0)=
[k\varphi_{b1} \,e^{kz}-k\varphi_{b2}\,e^{-kz}]_{k=0}
\nonumber\\
&=&\phi_0\,\frac{2\Lambda +z}{4\Lambda +s}\approx {\phi_0\over 2}\,.
\label{eq66}
\end{eqnarray}
Similarly, in the domain (c):
\begin{equation}
\Phi_c(z)=\phi_0\,\frac{2\Lambda -z}{4\Lambda +s}\approx
{\phi_0\over 2}\,.
\label{eq67}
\end{equation}
Therefore, $\Phi\approx\phi_0/2$ everywhere throughout the system.
\par
Since we have in this case only one pancake per the period $2s$ of
the structure, the line energy of the stack $\epsilon=\epsilon_p/2s$
where $ \epsilon_p$ is the energy per pancake:
\begin{eqnarray}
{8\pi\epsilon_p\over\phi_0}&=&\varphi_b(s)-\varphi_a(s)=
2\varphi_b(s)
\nonumber\\
&=&2\int\frac{d^2{\bf k}}{(2\pi)^2}
(\varphi_{b1}e^{ks}+\varphi_{b2}e^{-ks})\,.
\label{eq68}
\end{eqnarray}
\par
As above, within the integration domain we can simplify the denominator,
$D_2\approx 8k\Lambda$, and set $s=0$ in the integrand. Then we obtain
after integration within $2\pi/R, 2\pi/\xi$:
\begin{equation}
\epsilon =\Big(\frac{\phi_0 }{8\pi\lambda_{ab}}\Big)^2 \ln
{R\over \xi}\,.
\label{eq69}
\end{equation}
Comparing this with the line energy of a standard vortex
$(\phi_0/4\pi\lambda_{ab})^2\ln\kappa$ we see that dilute stacks
considered here are possible in small samples (whiskers) of a size
$R<\lambda_{ab}\kappa^3$.
\par
\section{Discussion}
We now estimate how the energy cost (\ref{eq43}) of the subsurface vortex
termination is affected by the Josephson coupling. Clearly, this coupling
breaks the cylindrical symmetry of the field associated with the straight
stack of pancake vortices, and the flux (\ref{eq39}) will not be spread 
even in all directions in the space between the top layer and the rest. 
Still, for small samples $R<\lambda_J$, in which the string cannot fully
develop, the asymmetry can be disregarded. Since the phase difference is
zero for perfectly aligned stack of pancakes for all pairs of layers
except the top one, the Josephson energy can be estimated as
\begin{eqnarray}
\epsilon_J
&=&\frac{\phi_0^2}{8\pi^3s\lambda_c^2}
\int_0^Rr\,dr \int_0^{2\pi}d\theta(1-\cos\theta)
\nonumber\\
&\approx&2s\Big(\frac{\phi_0R}{4\pi \lambda_{ab}\lambda_J}\Big)^2=
\frac{\phi_0^2}{8\pi^2\Lambda}\,{R^2\over\lambda_J^2}\,.
\label{eq70}
\end{eqnarray}
On the other hand, for $R\gg\lambda_J$, the integral in (\ref{eq70}) is
roughly proportional to the string area $R\lambda_J$ because $\lambda_J$
is an estimate for the string width:
\begin{equation}
\epsilon_J=\frac{\phi_0^2}{8\pi^3s\lambda_c^2}\,R\lambda_J
=\frac{\phi_0^2}{8\pi^3\Lambda}\,{R\over\lambda_J}\,.
\label{eq71}
\end{equation}
The energies (\ref{eq70})  or (\ref{eq71}) should be added to the
estimate (\ref{eq43}) for the  energy cost of the subsurface vortex
termination. \par
Taking, for example, Bi-2212 with $s=15\,$\AA, $\lambda_J=4500\,$\AA,
$\kappa =\lambda_{ab}/\xi_{ab}=50$, we have for $R=1\,\mu$:
\begin{equation}
\Delta\epsilon=
{\phi_0^2\over 16\pi^2\Lambda}
\Big[\ln\Big(\frac{R}{\kappa\lambda_{ab}(4\pi)^{3/2}}\,\Big)
+{2R\over \pi\lambda_J}\Big]\,.
\label{eq72}
\end{equation}
We estimate the logarithm here as $\approx -2$ (disregarding the factor
$(4\pi)^{-3/2}$  which could reduce this estimate to $-4$). The
Josephson correction is $2R/\pi\lambda_J\approx 1.4$. Thus, in
micron-size samples of Bi-2212, the termination of vortices under the
surface is possible.
\par
The above treatment is concerned with the case of zero applied
field. The energy cost of the subsurface termination should increase 
with increasing applied field. To have an estimate of this increase, one
can evaluate the work needed to remove a pancake from the top layer of a
cylindrical samle of radius $R$ in the presence of a small external field
$H_a$. Since one layer screening ability is weak, we assume the field
uniform, which corresponds to the asymuthal current density
$j=cA/4\pi\lambda^2=cH_a r/8\pi\lambda^2$ (the vector potential can be
taken as $A=H_ar/2$). The  work of the Lorentz force to remove a
pancake from the center is now readily obtained as   $\approx H_a
\phi_0R^2/16\pi\Lambda$. To compare with the estimate (\ref{eq72}) we
write the extra cost due to the applied field as
\begin{equation}
 \epsilon_a=
{\phi_0^2\over 16\pi^2\Lambda}\, {H_a\pi R^2\over \phi_0 } \,.
\label{eq73}
\end{equation}
Since the expression in parentheses of Eq. (\ref{eq72}) is on the order one,
we can say that the energy gain of the subsurface termination is  lost
in fields $H_a\sim\phi_0/\pi R^2$ which gives $H_a\sim 10\,$G for
$R=1\,\mu$. This sets an approximate upper bound on fields in which the
subsurface termination may occur.
\par
We would like to mention in conclusion that, since configurations
considered here and similar ones are not forbidden by topology, they
should be taken into account while studying fluctuations in layered
weakly coupled materials. They are certainly of interest for physics of
layered organic superconductors in which $\lambda_c$ may reach a
tens-of-micron size at low temperatures (see, e.g.,
Ref. \onlinecite{organic}) as well as for superconducting multilayers
where the Josephson coupling can be reduced by varying the interlayer
spacing.
\par
\acknowledgments
This research was supported by grant No. 96-00048 from the
United States - Israel Binational Science Foundation (BSF),
Jerusalem, Israel. The work at Ames was supported by the Office of Basic
Energy Sciences, U.S. Department of Energy. \par
\end{multicols}
\end{document}